\documentclass[intlimits,twoside,a4paper]{article}

\usepackage{amsmath,amssymb}
\usepackage{graphicx}
\usepackage{wrapfig}

\usepackage[T2A]{fontenc}
\usepackage[cp1251]{inputenc}

\usepackage{cmpj2}

\issue{2012}{15}{1}{13602}

\doinumber{10.5488/CMP.15.13602}

\title[Generalized equation of state applied to metals]%
{Generalized linear isotherm regularity equation \\ of state applied to metals}

 \author[H. Sun \textsl{et al.}]{H. Sun\refaddr{ad1}, J.X. Sun\refaddr{ad1,ad2}\thanks{E-mail: sjx@uestc.edu.cn}\,, W.J.~Yu\refaddr{ad1}, J.~Tang\refaddr{ad1}}

\addresses{
\addr{ad1} Department of Applied Physics, University of Electronic Science and Technology
of China, \\ Chengdu 610054, China
\addr{ad2} Laboratory for Shockwave and Detonation Physics, Southwest Institute of Fluid Physics,\\
Mianyang 621900, China
}

\date{Received September 14, 2011, in final form November 16, 2011}

\authorcopyright{H. Sun, J.X. Sun, W.J.~Yu, J.~Tang, 2012}

\begin{document}

\maketitle

\begin{abstract}
 A three-parameter equation of state (EOS) without physically
incorrect oscillations is proposed based on the generalized Lennard-Jones
(GLJ) potential and the approach in developing linear isotherm regularity
(LIR) EOS of Parsafar and Mason [J. Phys. Chem., 1994, \textbf{ 49}, 3049]. The proposed
(GLIR) EOS can include the LIR EOS therein as a special case. The
three-parameter GLIR, Parsafar and Mason (PM) [Phys. Rev. B, 1994, \textbf{ 49}, 3049],
Shanker, Singh and Kushwah (SSK) [Physica B, 1997, \textbf{ 229}, 419], Parsafar, Spohr
and Patey (PSP) [J. Phys. Chem. B, 2009, \textbf{ 113}, 11980], and reformulated PM and
SSK EOSs are applied to 30 metallic solids within wide pressure ranges. It is
shown that the PM, PMR and PSP EOSs for most solids, and the SSK and SSKR EOSs for
several solids, have physically incorrect turning points, and pressure
becomes negative at high enough pressure. The GLIR EOS is capable not only
of overcoming the problem existing in other five EOSs where the pressure becomes
negative at high pressure, but also gives results superior to other EOSs.

\pacs  64.30.Ef,  64.10.+h,  05.70.Ce
\end{abstract}

\section{ Introduction }

The equation of state (EOS) describes the relationships of a system among
thermodynamic variables such as pressure, temperature and volume, which plays
an important role in many fields, such as condensed-matter physics and
geophysics. The Murnaghan~\cite{1} and Birch~\cite{2,3} EOSs for solids are widely used
in geophysics. Since Rose et al.~\cite{4} proposed in 1986 that there exists a
universal EOS (UEOS) being valid for all types of solids through analyzing the
energy band data, a lot of forms of UEOS have been put forward~\cite{5,6,7,8,9,10,11,12,13,14,15,16}.
However, some of them have been applied to the study of thermodynamic properties of
liquids~\cite{6,7,8,9,10,11,12,13,14,15,16,17,18,19}, while the traditional Tait EOS has also been used as universal
equation both for solids~\cite{11,12} and liquids~\cite{13,14,15,16,17,18,19}.

In 1994, Parsafar and Mason (PM) proposed the following EOS by using a series
expansion of internal energy~\cite{11}
\begin{equation}
\label{eq:1}
P\left(  \frac{V}{V_{0}}\right)  ^{2}=C_{0}+C_{1}\left(  \frac{V_{0}}%
{V}\right)  +C_{2}\left(  \frac{V_{0}}{V}\right)  ^{2}.%
\end{equation}
Here, $V_{0}$ is the volume at zero pressure. $C_{0}$, $C_{1}$, $C_{2}$ are
three coefficients in the PM EOS. In 1997, Shanker, Singh and Kushwah (SSK)
proposed the following EOS~\cite{12,13}
\begin{equation}
\label{eq:2}
P=D_{0}+D_{1}\left(  \frac{V_{0}}{V}\right)  +D_{2}\left(  \frac{V_{0}}%
{V}\right)  ^{2},%
\end{equation}
where $D_{0}$, $D_{1}$, $D_{2}$ are three coefficients in the SSK EOS. It can be
seen that the SSK EOS can be expressed as volume-analytic and
pressure-analytic forms.

In 1994, Parsafar and Mason proposed the following linear isotherm regularity
(LIR) EOS for gases and liquids based on the Lennard-Jones (LJ) (12--6)
potential~\cite{14}
\begin{equation}
\label{eq:3}
\left(  Z-1\right)  \left(  \frac{V}{V_{0}}\right)  ^{2}=A_{0}+A_{2}\left(
\frac{V_{0}}{V}\right)  ^{2}.%
\end{equation}
Here, $Z$ is the compressibility factor, which is equal to $PV/RT$. The upper
density limit of LIR~\cite{14} is less certain but seems to be the freezing line
for liquids ($T<T_{\mathrm{c}}$) and at least about twice the Boyle density for
supercritical fluids. LIR EOS has been extended to mixtures~\cite{15} and to other
forms~\cite{16,17,18} through adopting different potential functions, including the
exponential-6~\cite{16}, LJ (6--3)~\cite{17}, LJ (885--4)~\cite{18}, and LJ (12--6--3)~\cite{19} potentials.

Recently, Parsafar, Spohr and Patey (PSP)~\cite{19}, extended the equation~(\ref{eq:3}) to the following
form with three parameters based on an effective near-neighbor pair
interaction of an LJ (12--6--3) potential
\begin{equation}
\label{eq:4}
\left(  Z-1\right)  \left(  \frac{V}{V_{0}}\right)  ^{2}=A_{0}+A_{1}\left(
\frac{V_{0}}{V}\right)  +A_{2}\left(  \frac{V_{0}}{V}\right)  ^{2}.%
\end{equation}
The PSP EOS can be equivalently reformulated as truncated Virial form
\begin{equation}
\label{eq:5}
P=\frac{RT}{V}+\frac{Q_{1}}{V^{2}}+\frac{Q_{2}}{V^{3}}+\frac{Q_{3}}{V^{5}}\,.%
\end{equation}
Parsafar et al.~\cite{19} claimed that the PSP EOS~(\ref{eq:4}) can be applied to all fluids and
solids, and their application for solids~\cite{19} does not reveal any pressure or
temperature limitations.

However, we noticed that the PM EOS~(\ref{eq:1}) and the PSP EOS~(\ref{eq:4}), (\ref{eq:5}) are physically
wrong at high pressure conditions for some solids. This is because the coefficients
$C_{2}$ in equation~(\ref{eq:1}) and $A_{2}$ in equation~(\ref{eq:4}) should be positive for all solids to
ensure a physically correct tendency at high pressure, $P\rightarrow\infty$ as $V\rightarrow0$.
However, the values of $C_{2}$ for most solids studied in this paper are
negative; and the values of $A_{2}$ for solids NaCl and CaO studied by
Parsafar et al.~\cite{19}, and for most solids studied in this paper are also
negative. This leads to an unphysical tendency, $P\rightarrow-\infty$ as $V\rightarrow0$.

The incorrect tendency makes the PM~\cite{11} and PSP~\cite{19} EOSs inapplicable to
high pressure conditions. We may preliminarily analyze the reason for
the failure of two EOSs as follows. Holzapfel~\cite{20} has pointed out that the limitation of
an EOS as the volume tends to zero, should be the Tomas--Fermi (TF) model, $P\propto V^{-5/3}$. The repulsion terms in PM~\cite{11} and PSP~\cite{19} EOSs are, $P\propto V^{-4}$ and $P\propto V^{-5}$, respectively. Their exponent numbers 4 and 5 are far larger than 5/3, and are too hard for solids. In order to fit experimental $P-V$ data at low and
middle pressure ranges, the optimized $C_{2}$ and $A_{3}$ should take on negative values.

In this work, we propose generalized LIR (GLIR) EOS based on a
near-neighbor pair potential of the extended Lennard-Jones ($m_{1}$,$n_{1}$)
type. The GLIR contains three parameters and can overcome the defect appearing
in the PM EOS~(\ref{eq:1}) and PSP EOS~(\ref{eq:4}). In section~2, the three-parameter GLIR EOS
is proposed. In section~3, equations~(\ref{eq:1}) and (\ref{eq:2}) and their modified version, PSP
EOS (\ref{eq:4}) and the GLIR EOS are applied to twenty solids within wide pressure
ranges of hundreds GPa and at ambient temperature, the results being analyzed
and discussed. In section~4, the conclusion is presented.

\section{Analytic equations of state}

We adopt the effective pair interaction of an extended Lennard-Jones ($m_{1}%
$,$n_{1}$) type potential~\cite{10,17,19,20}
\begin{equation}
\label{eq:6}
\varepsilon\left(  r\right)  =\frac{\varepsilon_{0}}{m_{1}-n_{1}}\left[
n_{1}\left(  \frac{r_\mathrm{e}}{r}\right)  ^{m_{1}}-m_{1}\left(  \frac{r_\mathrm{e}}%
{r}\right)  ^{n_{1}}\right].
\end{equation}
It is well known that the effective potentials for metals usually have
oscillating tails due to Friedel oscillations of electron density, and the
Lennard-Jones (LJ) potentials are not really appropriate for correct
reproduction of the energetics of metals. However, many works~\cite{10,11,12,13,14,15,16,17,18,19} have shown
that the LJ potentials can mimic many properties of metals in some compression
ranges. By adopting the nearest neighbor assumption~\cite{11}, the total
configurational energy of a solid is
\begin{eqnarray}
\label{eq:7}
U&=&\frac{N\delta\varepsilon_{0}}{2\left(  m_{1}-n_{1}\right)  }\left[
n_{1}\left(  \frac{r_\mathrm{e}}{r}\right)  ^{m_{1}}-m_{1}\left(  \frac{r_\mathrm{e}}%
{r}\right)  ^{n_{1}}\right]
\nonumber\\
&=&\frac{N\delta\varepsilon_{0}}{2\left(  m_{1}-n_{1}\right)  }\left[
n_{1}\left(  \frac{V_\mathrm{e}}{V_0}\right)  ^{m_{1}/3}-m_{1}\left(  \frac{V_\mathrm{e}}%
{V_0}\right)  ^{n_{1}/3}\right],
\end{eqnarray}
where $V=a^{3}/\gamma$, $V_{0}=\left(  r_\mathrm{e}\right)  ^{3}/\gamma$, $a$ is the nearest neighbor distance, and
$\delta$ is the mean coordination number~\cite{10,19}. Following Parsafar and Mason~\cite{8}, the
internal pressure can be obtained by the derivative of equation~(\ref{eq:6})
\begin{equation}
\label{eq:8}
P_\mathrm{int}=-\frac{\partial U}{\partial V}=\frac{m_{1}n_{1}N\delta\varepsilon_{0}%
}{6\left(  m_{1}-n_{1}\right)  }\left[  \left(  \frac{V_\mathrm{e}}{V_{0}}\right)
^{m_{1}/3+1}-\left(  \frac{V_\mathrm{e}}{V_{0}}\right)  ^{n_{1}/3+1}\right].
\end{equation}
Let us substitute the equation~(\ref{eq:7}) into the following internal energy equation
\begin{equation}
\label{eq:9}
P=T\left(  \frac{\partial P}{\partial T}\right)  _{V}-\left(  \frac{\partial
U}{\partial V}\right)  _{T}\,.%
\end{equation}
After integration, we derive the equation
\begin{equation}
\label{eq:10}
P=\frac{RT}{V}+A_{1}\left(  \frac{V_{0}}{V}\right)  ^{n_{1}/3+1}+A_{2}\left(
\frac{V_{0}}{V}\right)  ^{m_{1}/3+1}.%
\end{equation}
Here, $A_{1}$ and $A_{2}$ are functions of temperature.

In order to obtain an extended LIR EOS, we would limit parameters $m_{1}$ and
$n_{1}$ to satisfy the relationship, $m_{1}$ = 2 $n_{1}$, and
\begin{equation}
\label{eq:11}
m_{1}/3=2m,\qquad  n_{1}/3=m.
\end{equation}
Then, equation~(\ref{eq:7}) changes to the following form
\begin{equation}
\label{eq:12}
P=\frac{RT}{V}+A_{1}\left(  \frac{V_{0}}{V}\right)  ^{m+1}+A_{2}\left(
\frac{V_{0}}{V}\right)  ^{2m+1}.%
\end{equation}
By using  definition of compressibility, $Z=PV/RT$, equation~(\ref{eq:12}) can be reformulated following the generalized LIR (GLIR) EOS
\begin{equation}
\label{eq:13}
\left(  Z-1\right)  \left(  \frac{V}{V_{0}}\right)  ^{m}=B_{0}+B_{1}\left(
\frac{V_{0}}{V}\right)  ^{m},
\end{equation}
\begin{equation}
\label{eq:14}
B_{0}=\frac{A_{0}V_{0}}{RT}\,,\qquad B_{1}=\frac{A_{2}V_{0}}{RT}\,.
\end{equation}

It can be seen that the LIR EOS in equation~(\ref{eq:3}) can be included in the GLIR EOS~(\ref{eq:13}) as a special case when $m$ = 2. Since $m$ = 1, equation~(\ref{eq:13}) just reduces to
the virial EOS. Although the parameter number of PSP EOS~(\ref{eq:4}) is the same as
the three-parameter GLIR EOS~(\ref{eq:13}), equation~(\ref{eq:13}) with adjustable parameter $m$ is
more flexible and more accurate than equation~(\ref{eq:4}).

Otherwise, we found in our calculations that the PM~\cite{11} and SSK~\cite{12,13} EOSs
can be reformulated in the following forms:
\begin{eqnarray}
\label{eq:15}
P\left(  \frac{V}{V_{0}}\right)  ^{4}&=&C_{2}+C_{1}\left(  \frac{V}{V_{0}%
}\right)  +C_{0}\left(  \frac{V}{V_{0}}\right)  ^{2},\\
\label{eq:16}
P\left(  \frac{V}{V_{0}}\right)  ^{2}&=&D_{2}+D_{1}\left(  \frac{V}{V_{0}%
}\right)  +D_{0}\left(  \frac{V}{V_{0}}\right)  ^{2}.%
\end{eqnarray}
We name this form as PMR and SSKR EOSs. Although the PMR EOS and SSKR EOSs are
mathematically equivalent to the PM and SSK EOSs, they  physically
differ  from each other. This is because all of equations~(\ref{eq:1}), (\ref{eq:2}) and equations~(\ref{eq:15}), (\ref{eq:16}) can be
seen as Taylor expansion, but the expansion variable of equations~(\ref{eq:1}), (\ref{eq:2}) is ($V_{0}$/$V$), and that of equations~(\ref{eq:15}), (\ref{eq:16}) is ($V$/$V_{0}$). At zero pressure, both values of ($V_{0}$/$V$) and ($V$/$V_{0}$) are equal to 1. At high pressure, the values of ($V$/$V_{0}$) are smaller than 1, the Taylor expansions in equations~(\ref{eq:15}), (\ref{eq:16}) are fast convergent. However, the values of ($V_{0}$/$V$) are larger than 1 at high pressure, the Taylor expansions in equations~(\ref{eq:1}), (\ref{eq:2}) are slowly
convergent. Thus, the PMR and SSKR EOSs in equations~(\ref{eq:15}), (\ref{eq:16}) are more accurate than
the original PM and SSK EOSs in equations~(\ref{eq:1}), (\ref{eq:2}).

\section{Results and discussion}

Now we apply six EOSs to 28 metallic solids, including GLIR~(\ref{eq:13}), PM~\cite{11},
PMR~(\ref{eq:15}), SSK~\cite{12,13}, SSKR~(\ref{eq:16}) and PSP~\cite{19} EOSs. All experimental data are
taken from Kennedy and Keeler (1972)~\cite{21}, except for W~\cite{22}.

\begin{table}[ht]
\caption{The experimental data of $V_0$ (cm$^{3}$/mol) and comparison of average
relative errors ($\Delta_{\mathrm{p}}$\%) of pressure for 30 metallic solids calculated
from the GLIR, PM, PMR, SSK, SSKR, and PSP EOSs.}
\vspace{2ex}
\begin{center}
\small{
\begin{tabular}
[c]{|c|c|c|c|c|c|c|c|}
\hline
&  & GLIR & PM & PMR & SSK & SSKR & PSP\\
& $V_{0}$ & $\Delta_{\mathrm{p}}\%$ & $\Delta_{\mathrm{p}}\%$ & $\Delta_{\mathrm{p}}\%$ & $\Delta_{\mathrm{p}}\%$
& $\Delta_{\mathrm{p}}\%$ & $\Delta_{\mathrm{p}}\%$\\
\hline\hline
Cu & 7.115 & 0.54 & 0.54 & 0.50 & 9.34 & 4.65 & 0.43\\\hline
Mo & 9.387 & 0.80 & 1.99 & 1.45 & 1.08 & 1.05 & 1.32\\\hline
Zn & 9.166 & 0.30 & 0.55 & 0.39 & 9.34 & 5.46 & 0.48\\\hline
Ag & 10.27 & 0.38 & 0.47 & 0.41 & 6.22 & 4.10 & 0.45\\\hline
Pt & 9.098 & 0.70 & 0.70 & 0.70 & 2.15 & 1.69 & 0.70\\\hline
Ti & 12.01 & 0.68 & 3.38 & 2.03 & 2.07 & 1.07 & 2.09\\\hline
Ta & 10.80 & 0.66 & 1.14 & 0.86 & 0.68 & 0.64 & 0.89\\\hline
Au & 10.22 & 0.64 & 0.64 & 0.64 & 2.44 & 1.80 & 0.64\\\hline
Pd & 8.896 & 0.72 & 0.72 & 0.72 & 2.21 & 1.55 & 0.72\\\hline
Zr & 14.02 & 0.62 & 7.02 & 4.21 & 3.70 & 2.13 & 3.51\\\hline
Cr & 7.231 & 0.99 & 1.00 & 1.00 & 1.35 & 1.20 & 1.00\\\hline
Co & 6.689 & 0.65 & 0.61 & 0.61 & 0.60 & 0.60 & 0.60\\\hline
Ni & 6.592 & 0.60 & 0.61 & 0.61 & 1.13 & 0.98 & 0.62\\\hline
Nb & 10.83 & 1.71 & 1.90 & 2.00 & 2.15 & 2.20 & 1.76\\\hline
Cd & 13.00 & 0.24 & 0.30 & 0.29 & 4.08 & 2.92 & 0.31\\\hline
Al & 10.00 & 0.85 & 0.76 & 0.76 & 0.87 & 0.66 & 0.49\\\hline
Th & 19.97 & 0.36 & 0.95 & 0.69 & 0.78 & 0.53 & 0.86\\\hline
V & 8.365 & 0.39 & 0.70 & 0.58 & 0.41 & 0.39 & 0.49\\\hline
In & 15.73 & 0.57 & 0.76 & 0.62 & 3.98 & 2.91 & 0.60\\\hline
Be & 4.890 & 0.43 & 0.64 & 0.55 & 0.61 & 0.52 & 0.51\\\hline
Pb & 18.27 & 0.30 & 0.31 & 0.31 & 2.86 & 2.06 & 2.09\\\hline
Sn & 16.32 & 0.26 & 0.31 & 0.29 & 2.50 & 1.81 & 0.29\\\hline
Mg & 14.00 & 0.33 & 0.59 & 0.47 & 0.27 & 0.26 & 0.66\\\hline
Ca & 26.13 & 0.61 & 5.67 & 4.05 & 2.91 & 1.69 & 4.98\\\hline
Tl & 17.23 & 0.29 & 0.29 & 0.29 & 1.38 & 1.08 & 0.28\\\hline
Na & 23.71 & 0.48 & 0.94 & 0.68 & 0.23 & 0.21 & 1.15\\\hline
K & 45.62 & 0.39 & 1.30 & 1.20 & 0.85 & 0.46 & 1.80\\\hline
Rb & 56.08 & 0.43 & 1.21 & 1.15 & 0.64 & 0.36 & 1.75\\
\hline\multicolumn{2}{|c|}{mean error} & 0.57 & 1.29 & 1.00 & 2.39 & 1.61 & 1.12\\
\hline
\end{tabular}}
\end{center}
\label{tab1}
\end{table}
In table~\ref{tab1}, we
list the volume at zero pressure $V_{0}$, average fitting errors of pressure
for the 28 solids. It can be seen that the GLIR~(\ref{eq:13}) yields the smallest
fitting errors for 20 solids, and for the other 8 solids the errors are also fairly
small. The fitting precision for different solids is fairly stable for the
GLIR EOS~(\ref{eq:13}), while instable for the other five EOSs. The largest errors among the 28 solids for the six EOSs are 1.71\% of Nb, 7.02\% of Zr, 4.21\% of Zr, 9.34\%
of Zn, 5.46\% of Zn, 4.98\% of Ca, respectively. In the last line of the table, we
list the total average error for the 28 solids. It can be seen that the GLIR
EOS yields the best results with average error 0.57\%; the PMR EOS yields second
best results with average error 1.00\%; the PSP EOS, PM EOS, SSKR EOS, and SSK
EOS subsequently give worse results with average errors 1.12\%, 1.29\%, 1.61\%
and 2.39\%, respectively.

\begin{table}[ht]
\caption{Optimized values of coefficients for the GLIR, PM and PMR EOSs
determined by fitting experimental compression data. The parameters for the GLIR EOS are dimensionless; and all parameters for PM and PMR EOSs are in GPa.}
\vspace{2ex}
\begin{center}
\small{
\begin{tabular}
[c]{|c|c|c|c|c|c|c|c|c|c|}
\hline
& \multicolumn{3}{|c|}{GLIR} & \multicolumn{3}{|c|}{PM} & \multicolumn{3}{|c|}{PMR}%
\\
\hline
& $m$ & $B_{0}$ & $B_{1}$ & $C_{0}$ & $C_{1}$ & $C_{2}$ & $C_{0}$ & $C_{1}$ &
$C_{2}$\\
\hline\hline
Cu & 0.906 & --449.53 & 448.57 & --153.86 & 167.29 & --13.44 & --13.24 & 166.74 &
--153.51\\\hline
Mo & 0.592 & --1731.45 & 1732.80 & --395.31 & 458.15 & --98.61 & --101.51 &
465.36 & --363.72\\\hline
Zn & 1.199 & --188.20 & 187.21 & --48.41 & 34.78 & 13.53 & 14.11 & 33.18 &
--47.34\\\hline
Ag & 1.197 & --368.37 & 367.35 & --81.38 & 55.72 & 25.60 & 26.14 & 54.37 &
--80.54\\\hline
Pt & 1.031 & --1000.33 & 999.46 & --272.52 & 263.83 & 8.69 & 9.54 & 261.87 &
--271.40\\\hline
Ti & 0.416 & --1185.34 & 1184.03 & --126.25 & 164.13 & --37.40 & --40.20 &
171.78 & --131.30\\\hline
Ta & 0.527 & --1655.67 & 1654.63 & --218.00 & 369.91 & --88.75 & --90.88 &
375.05 & --284.06\\\hline
Au & 1.004 & --762.52 & 761.51 & --184.79 & 183.79 & 0.70 & 1.52 & 181.85 &
--183.36\\\hline
Pd & 1.031 & --683.16 & 682.19 & --188.56 & 181.27 & 7.29 & 7.11 & 181.68 &
188.80\\\hline
Zr & 0.197 & --2763.5 & 2762.6 & --122.61 & 166.63 & --43.34 & --48.00 & 179.09 &
--130.68\\\hline
Cr & 0.924 & --609.57 & 608.36 & --209.26 & 226.66 & --17.46 & --17.58 & 226.93 &
--209.41\\\hline
Co & 0.729 & --733.72 & 732.70 & --257.16 & 318.67 & --61.49 & --61.29 & 318.20 &
--256.89\\\hline
Ni & 0.893 & --565.39 & 564.35 & --211.11 & 234.21 & --23.09 & --22.88 & 233.72 &
--210.83\\\hline
Nb & 0.582 & --1282.5 & 1281.3 & --250.03 & 331.05 & --81.09 & --82.18 & 333.60 &
--251.51\\\hline
Cd & 1.223 & --216.51 & 215.61 & --36.99 & 23.21 & 13.76 & 13.84 & 23.00 &
--36.86\\\hline
Al & 0.719 & --452.88 & 451.48 & --106.33 & 132.17 & --26.01 & --25.01 & 129.71 &
--104.84\\\hline
Th & 0.613 & --701.54 & 700.18 & --65.47 & 81.33 & --15.71 & --16.41 & 83.24 &
--66.73\\\hline
V & 0.569 & --939.67 & 938.80 & --223.73 & 293.70 & --69.84 & --70.98 & 296.38 &
--225.29\\\hline
In & 1.058 & --240.83 & 239.89 & --38.61 & 36.62 & 1.930 & 2.430 & 35.31 &
--37.75\\\hline
Be & 0.477 & --500.16 & 499.19 & --177.73 & 239.41 & --61.58 & --62.52 & 241.66 &
--179.05\\\hline
Pb & 1.022 & --323.72 & 322.61 & --43.61 & 42.49 & 1.090 & 1.240 & 42.12 &
--44.37\\\hline
Sn & 1.118 & --256.62 & 255.80 & --37.87 & 32.02 & 5.850 & 5.970 & 31.72 &
--37.69\\\hline
Mg & 0.592 & --338.70 & 337.60 & --44.32 & 55.83 & --11.41 & --11.81 & 56.91 &
--45.03\\\hline
Ca & 0.076 & --2812.2 & 2810.9 & --20.35 & 27.60 & --6.810 & --8.040 & 31.34 &
--23.05\\\hline
Tl & 1.154 & --216.85 & 215.66 & --28.85 & 21.85 & 6.990 & 7.160 & 21.43 &
--28.60\\\hline
Na & 0.540 & --112.14 & 111.09 & --6.740 & 8.370 & --1.540 & --1.660 & 8.740 &
--7.030\\\hline
K & 0.419 & --147.58 & 146.11 & --2.900 & 3.670 & --0.670 & --0.750 & 3.980 &
--3.160\\\hline
Rb & 0.457 & --115.00 & 113.29 & --1.780 & 2.230 & --0.360 & 2.440 & --1.970 &
1.210\\
\hline
\end{tabular}}
\end{center}
\label{tab2}
\end{table}
In tables~\ref{tab2} and~\ref{tab3}, we list the fitted parameters for the six EOSs, table~\ref{tab2}
shows that the values of $m$ in the GLIR EOS~(\ref{eq:13}) are smaller than 1 for 19
solids, and slightly larger than 1 for 10 solids. This implies that the
interactions in the metals are far softer than the LJ (12--6) potential, and
are approximately approaching the LJ (6--3) potential for the 10 solids, and
even softer than the LJ (6--3) potential for other 20 solids. The table
%
%
also
shows that the parameter $B_{1}$ in the GLIR EOS~(\ref{eq:13}) always takes on positive
values, and this ensures a correct tendency as the volume tends to infinity.

\begin{figure}[p]
\begin{minipage}{0.45\textwidth}
\centerline{ \includegraphics[width=\textwidth]{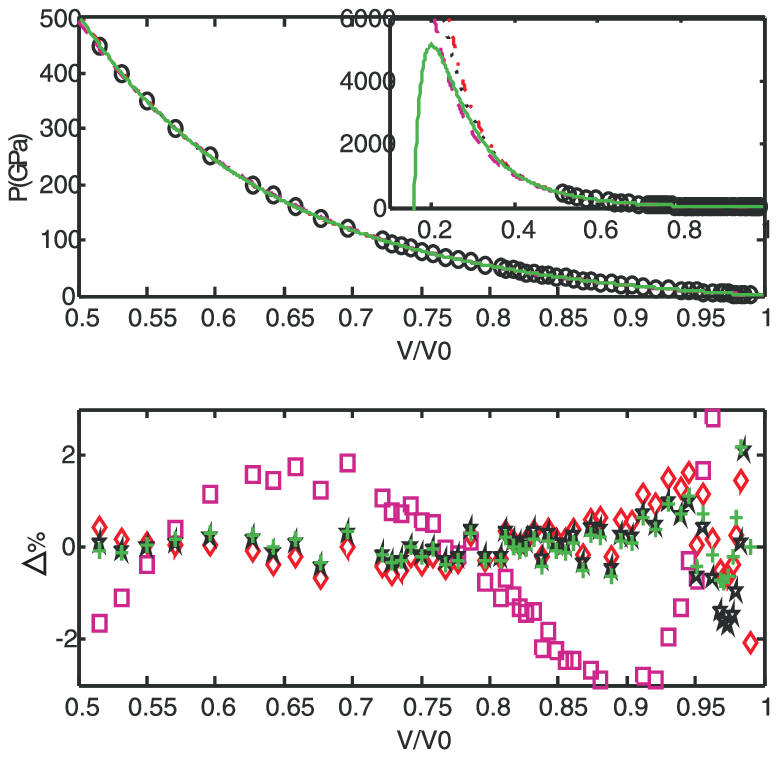}}%
\end{minipage}
\hfill
\begin{minipage}{0.45\textwidth}
\centerline{
\includegraphics[width=\textwidth]{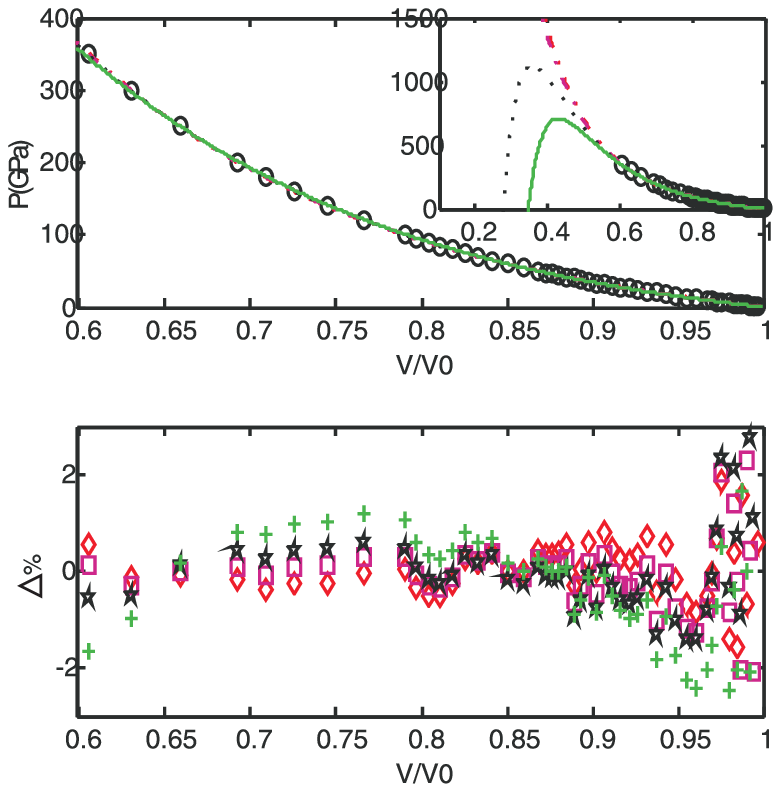}}%
\end{minipage}
\\%
\parbox[t]{0.45\textwidth}{%
\centerline{(a)}%
}%
\hfill%
\parbox[t]{0.45\textwidth}{%
\centerline{(b)}%
}%
\\[2ex]
\begin{minipage}{0.45\textwidth}
\centerline{ \includegraphics[width=\textwidth]{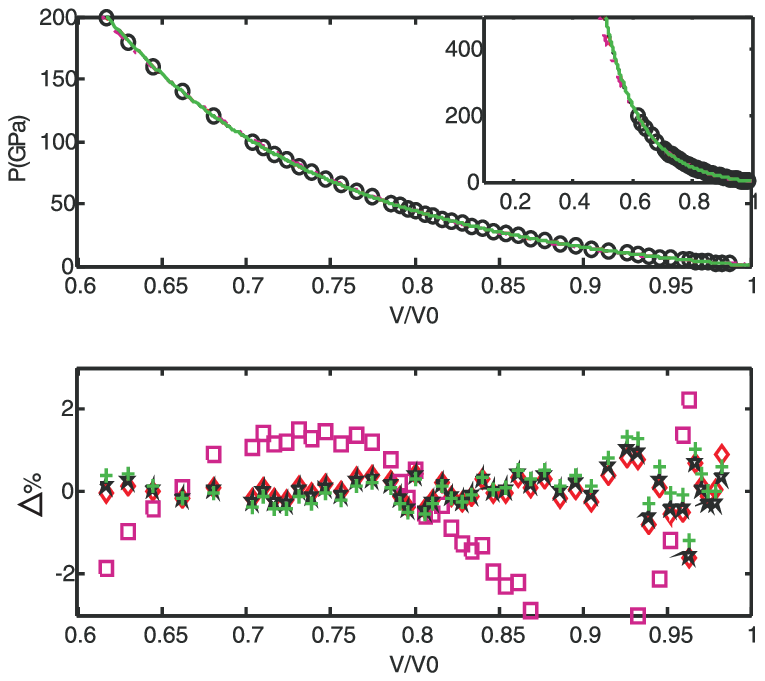}}%
\end{minipage}
\hfill
\begin{minipage}{0.45\textwidth}
\centerline{
\includegraphics[width=\textwidth]{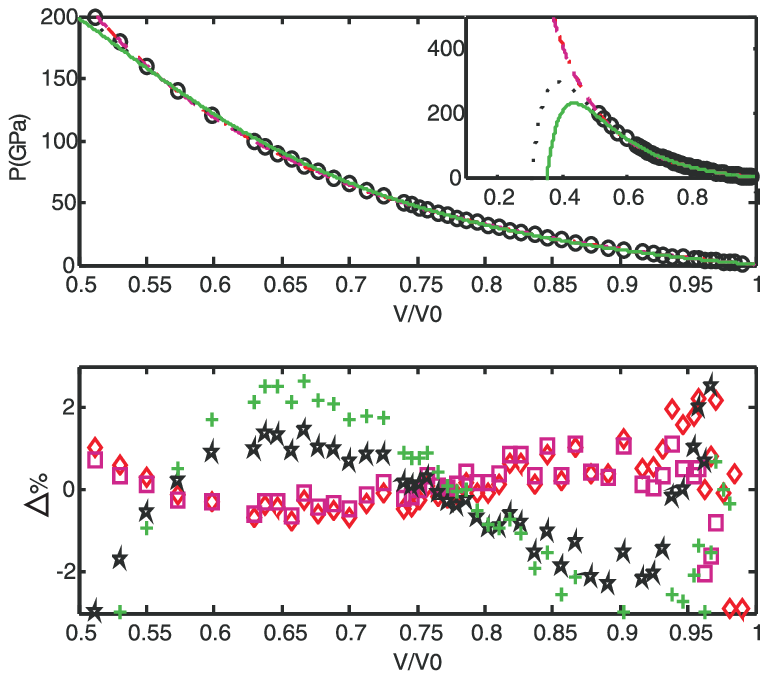}}%
\end{minipage}
\\%
\parbox[t]{0.45\textwidth}{%
\centerline{(c)}%
}%
\hfill%
\parbox[t]{0.45\textwidth}{%
\centerline{(d)}%
}%
\\[2ex]
\begin{minipage}{0.45\textwidth}
\centerline{
\includegraphics[width=\textwidth]{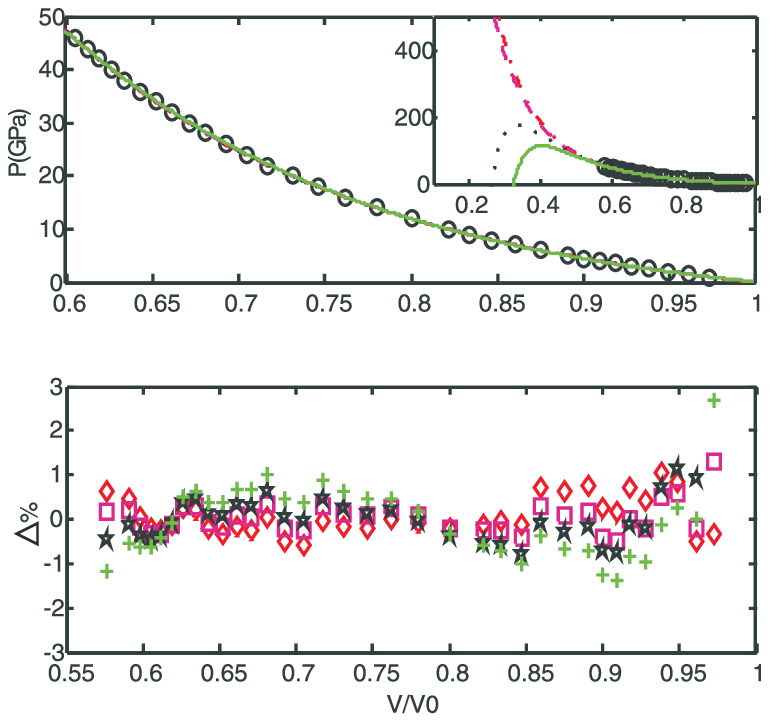}}%
\end{minipage}
\hfill
\begin{minipage}{0.45\textwidth}
\centerline{
\includegraphics[width=\textwidth]{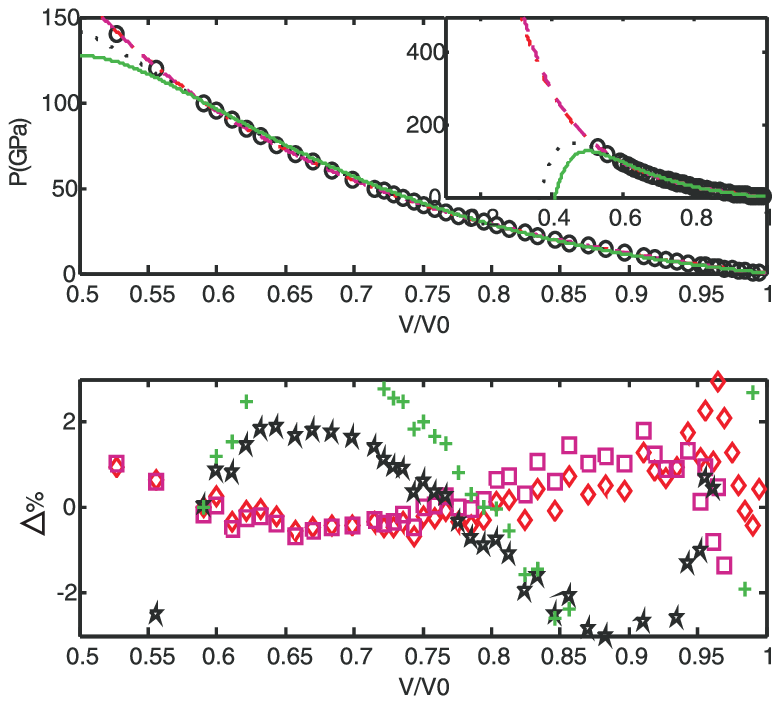}}%
\end{minipage}
\\%
\parbox[t]{0.45\textwidth}{%
\centerline{(e)}%
}%
\hfill%
\parbox[t]{0.45\textwidth}{%
\centerline{(f)}%
}%
\caption{(Color online) Comparison of compression curves of Cu (a), Mo (b), Ag (c), Ti (d), Mg (e), and Zr (f) calculated by using different
equations with experimental data ($\circ$): solid line, PSP EOS; dashed
line, SSKR EOS; dot line, PMR EOS; dot-dashed line, GLIR EOS. And comparison
of percentage error of pressure calculated using different equations:
$+$, PSP EOS; $\square$, SSKR EOS; $\star$, PMR EOS; $\Diamond$, GLIR EOS.}
\label{fig1}
\end{figure}

\begin{figure}[!h]
\vspace{-2mm}
\begin{minipage}{0.45\textwidth}
\centerline{
\includegraphics[width=\textwidth]{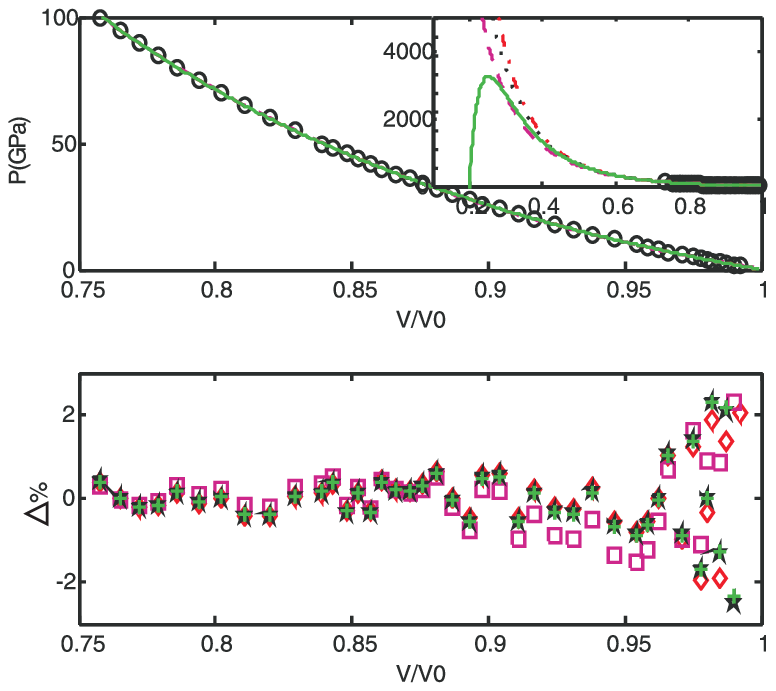}}
\end{minipage}
\hfill
\begin{minipage}{0.45\textwidth}
\centerline{
\includegraphics[width=\textwidth]{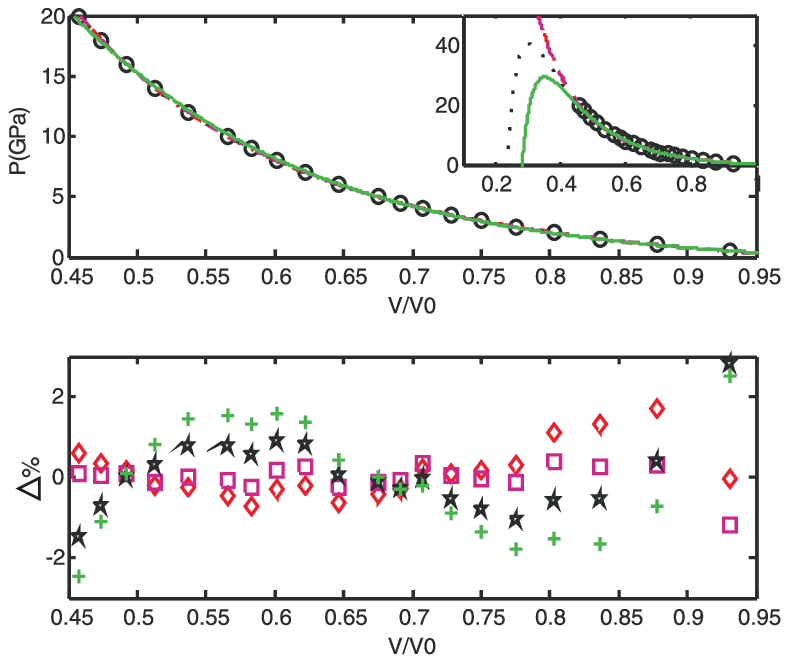}}
\end{minipage}
\\%
\parbox[t]{0.45\textwidth}{%
\centerline{(a)}%
}%
\hfill%
\parbox[t]{0.45\textwidth}{%
\centerline{(b)}%
}%
\caption{(Color online) Comparison of compression curves of Ni (a) and Na (b) calculated by using different
equations with experimental data ($\circ$): solid line, PSP EOS; dashed
line, SSKR EOS; dot line, PMR EOS; dot-dashed line, GLIR EOS. And comparison
of percentage error of pressure calculated using different equations:
$+$, PSP EOS; $\square$, SSKR EOS; $\star$, PMR EOS; $\Diamond$, GLIR EOS.}
\label{fig2}
\vspace{-2mm}
\end{figure}

\begin{figure}[ht]
\begin{minipage}{0.47\textwidth}
\centerline{
\includegraphics[width=\textwidth]{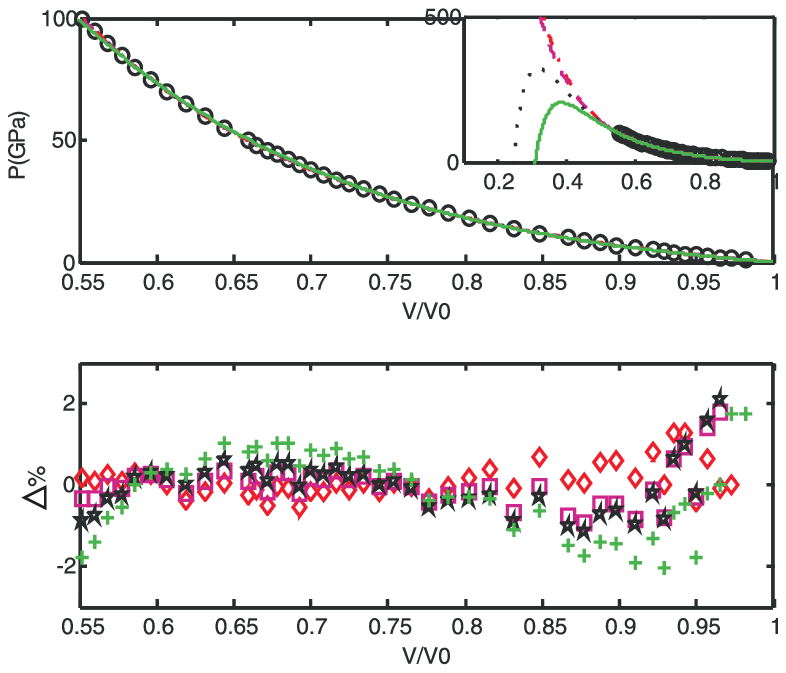}}
\end{minipage}
\hfill
\begin{minipage}{0.42\textwidth}
\centerline{
\includegraphics[width=\textwidth]{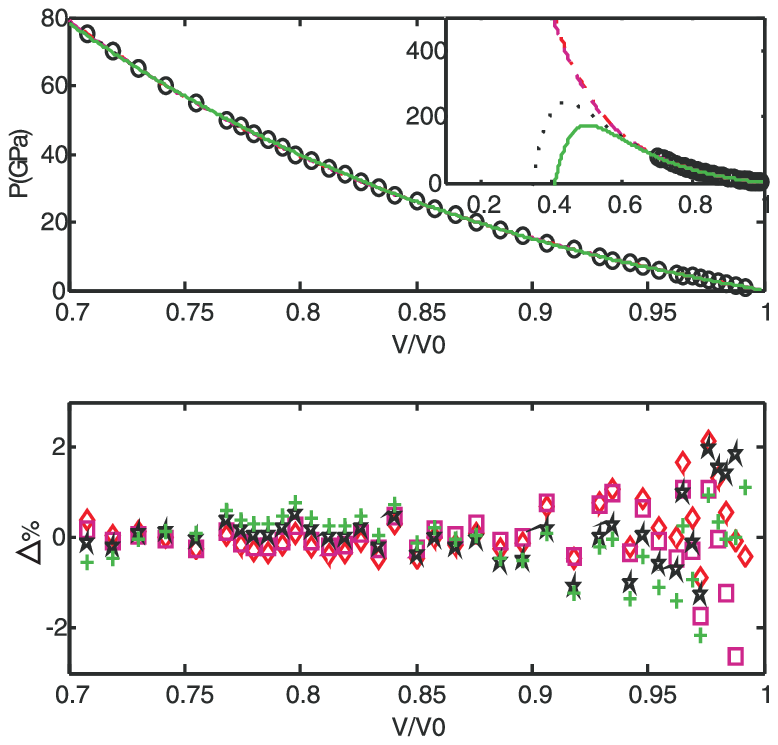}}
\end{minipage}
\\%
\parbox[t]{0.47\textwidth}{%
\centerline{(a)}%
}%
\hfill%
\parbox[t]{0.42\textwidth}{%
\centerline{(b)}%
}%
\caption{(Color online) Comparison of compression curves of Th (a) and Be (b)  calculated by using different
equations with experimental data ($\circ$): solid line, PSP EOS; dashed
line, SSKR EOS; dot line, PMR EOS; dot-dashed line, GLIR EOS. And comparison
of percentage error of pressure calculated using different equations:
$+$, PSP EOS; $\square$, SSKR EOS; $\star$, PMR EOS; $\Diamond$, GLIR EOS.}
\label{fig3}
\end{figure}

However, the values of $C_{2}$ in the PM and PMR EOSs are negative for 18 and
25 solids, respectively. The values of $D_{2}$ in the SSKR EOS, $A_{2}$ in the
PSP EOS are also negative for 2 and 18 solids, respectively. For these solids,
the corresponding EOSs may exhibit a physically incorrect tendency as the volume
tends to infinity. To compare, the GLIR EOS~(\ref{eq:13}) is not only the most
precise one, but also is a unique EOS that does not exhibit a physically
incorrect tendency among the six EOSs studied in this work.
In figures~\ref{fig1}, \ref{fig2} and~\ref{fig3}, we plot the experimental compression data and the curves calculated
using the GLIR, PMR, SSKR and PSP for 10 solids, including Cu, Mo, Ag, Ti,
Ta, Zr, Ni, Nb, Th and Be. These figures show that the calculated compression
curves from the GLIR and SSKR EOSs are correct at high pressure for the 10
solids, although for Zr, the parameter $D_{2}$ in the SSKR EOS takes on a negative
value. But the PMR EOS yields incorrect compression curves at high pressure for
7 of 10 solids, except for the solid Cu, Ag, and Ni. And the turn point is in
the range $V$/$V_{0}\approx$ ($0.3\div0.5$) for the 7 solids. Moreover, the
PSP EOS also yields incorrect compression curves at high pressure for 9 of 10
solids, except for the solid Ag. And the turn point is about $V$/$V_{0}$
$\approx$ 0.3 for solids Cu and Ni; about $V$/$V_{0}\approx$ 0.5 for other 7 solids.

\begin{table}[ht]
\caption{Optimized values of coefficients for the SSK, SSKR and PSP EOSs
determined by fitting experimental compression data. The parameters for the PSP EOS are dimensionless; and all parameters for SSK and SSKR EOSs are in GPa.}
\vspace{2ex}
\begin{center}
\small{
\renewcommand\baselinestretch{0.89}\normalsize
\begin{tabular}
[c]{|c|c|c|c|c|c|c|c|c|c|}
\hline
& \multicolumn{3}{|c|}{SSK} & \multicolumn{3}{|c|}{SSKR} & \multicolumn{3}{|c|}{PSP}%
\\
\hline
& $D_{0}$ & $D_{1}$ & $D_{2}$ & $D_{0}$ & $D_{1}$ & $D_{2}$ & $A_{0}$ &
$A_{1}$ & $A_{2}$\\
\hline\hline
Cu & 293.73 & --686.40 & 394.71 & 381.53 & --650.37 & 269.96 & 427.41 &
--419.20 & --9.020\\\hline
Mo & 156.70 & --578.16 & 421.54 & 421.11 & --577.10 & 156.06 & 1316.8 &
--1209.8 & --107.85\\\hline
Zn & 260.70 & --528.62 & 271.29 & 256.94 & --489.17 & 234.34 & 192.99 &
--206.99 & 12.750\\\hline
Ag & 326.55 & --720.31 & 395.34 & 382.78 & --688.72 & 307.02 & 359.42 &
--389.56 & 28.99\\\hline
Pt & 466.76 & --1193.8 & 727.49 & 720.07 & --1176.7 & 456.98 & 996.17 &
--1008.9 & 11.81\\\hline
Ti & 8.260 & --121.04 & 112.44 & 114.63 & --127.03 & 12.22 & 585.21 & --536.06 &
--49.24\\\hline
Ta & 72.99 & --344.84 & 271.82 & 272.46 & --346.38 & 73.90 & 1179.1 & --1066.7 &
--113.03\\\hline
Au & 325.64 & --816.00 & 490.93 & 484.10 & --799.81 & 316.14 & 761.26 &
--763.39 & 1.120\\\hline
Pd & 344.40 & --865.63 & 521.72 & 513.51 & --846.47 & 333.32 & 682.39 &
--690.84 & 7.450\\\hline
Zr & --40.14 & --22.08 & 61.86 & 64.60 & --29.40 & --35.39 & 690.45 & --617.45 &
--73.00\\\hline
Cr & 259.25 & --700.19 & 441.11 & 437.08 & --609.92 & 253.56 & 605.12 &
--591.28 & --15.00\\\hline
Co & 161.67 & --516.19 & 354.61 & 354.11 & --515.05 & 161.00 & 663.78 &
--618.34 & --46.25\\\hline
Ni & 241.20 & --661.39 & 420.41 & 417.51 & --654.68 & 237.36 & 553.90 &
--537.02 & --17.83\\\hline
Nb & 70.990 & --313.58 & 242.46 & 241.83 & --312.12 & 70.140 & 1029.9 &
--931.42 & --99.57\\\hline
Cd & 175.83 & --376.06 & 201.58 & 195.08 & --359.21 & 165.11 & 211.34 &
--231.68 & 19.28\\\hline
Al & 64.27 & --207.75 & 143.33 & 144.62 & --210.93 & 66.190 & 39.50 & --370.24 &
--26.01\\\hline
Th & 39.87 & --129.44 & 89.70 & 89.03 & --127.60 & 38.66 & 500.66 & --467.25 &
--33.83\\\hline
V & 72.94 & --302.04 & 229.17 & 228.67 & --300.87 & 72.25 & 733.25 & --663.29 &
--70.70\\\hline
In & 109.23 & --241.91 & 133.57 & 129.82 & --231.91 & 102.73 & 244.77 &
--249.76 & 3.890\\\hline
Be & 26.94 & --175.10 & 148.11 & 148.97 & --177.14 & 28.14 & 338.68 & --303.43 &
--36.13\\\hline
Pb & 103.88 & --239.29 & 136.04 & 132.39 & --229.89 & 97.94 & 276.25 & --287.00 &
11.52\\\hline
Sn & 110.10 & --251.61 & 142.07 & 138.48 & --242.59 & 104.52 & 256.41 &
--268.25 & 10.89\\\hline
Mg & 21.76 & --77.32 & 55.60 & 55.56 & --77.22 & 21.69 & 238.92 & --222.07 &
--17.46\\\hline
Ca & --14.73 & 6.430 & 8.020 & 8.840 & 3.940 & --12.93 & 220.27 & --198.94 &
--20.39\\\hline
Tl & 85.87 & --200.18 & 114.59 & 112.74 & --195.66 & 83.15 & 210.86 & --225.44 &
13.47\\\hline
Na & 3.410 & --12.71 & 9.300 & 9.330 & --12.81 & 3.480 & 61.28 & --58.10 &
--3.550\\\hline
K & 0.170 & --3.950 & 3.700 & 3.810 & --4.330 & 0.490 & 48.05 & --44.26 &
--2.410\\\hline
Rb & 0.530 & --3.450 & 2.850 & 2.910 & --3.680 & 0.750 & 35.90 & --32.56 & --1.400\\\hline
\end{tabular}}
\renewcommand\baselinestretch{1}\normalsize
\end{center}
\label{tab3}
\end{table}

In these figures, we also plot the variation of relative errors of pressure
with compression ratio $V$/$V_{0}$. It can be seen from these figures that,
for solids Cu and Ag, the oscillations of relative errors from the SSKR EOS
are the most prominent, and are the same from other three EOSs; for solids
Ti and Zr, the oscillations of relative errors from the PSP EOS and PMR EOS
are more evident than the SSKR and GLIR EOSs; and for other solids, the
oscillations from all four EOSs are equivalent with each other. It is
notable that the relative errors from the GLIR EOS are most stable and fairly
small for all 10 solids and for all compression ratio ranges. These results
show that the GLIR EOS can be seen as the best one among six EOSs studied in
this work.

\section{Conclusion}

In conclusion, we develop a three-parameter GLIR based on the GLJ potential
and the approach of Parsafar and Mason~\cite{14} in developing the LIR EOS. Comparing with other five EOSs popular in literature, the precision of the
GLIR EOS developed in this paper is superior to other EOSs. The GLIR EOS is capable of overcoming the problem existing in other EOSs where the pressure becomes negative
at high enough pressure conditions.

\section*{Acknowledgements}

This work is supported by the Joint Fund of NSFC and CAEP under Grant No.
10876008, and by the Innovation Fund of UESTC under Grant No. 23601008.

\ukrainianpart

\title{Узагальнене рівняння стану, застосовне до металів}

\author{Г. Сан\refaddr{ad1}, Дж.Г. Сан\refaddr{ad1,ad2}, В.Дж. Йу\refaddr{ad1}, Дж. Танг\refaddr{ad1}}
\addresses{\addr{ad1}Кафедра прикладної фізики, Китайський університет електроніки та технологій,  Ченду 610054, КНР
\addr{ad2} Лабораторія фізики ударної хвилі і детонації, Південно-Західний інститут фізики плинів,\\
Міанян 621900, КНР
}

\makeukrtitle

\begin{abstract}
\tolerance=3000%
Запропоновано трипараметричне рівняння стану без фізично некоректних осциляцій, що базується на узагальненому потенціалі
Леннарда-Джонса (GLJ) і підході Парсафара і Мейсона [Parsafar~G.A., Mason~E.A.,
J.~Phys. Chem.,  1994, \textbf{49}, 3049] до виведення рівняння стану з регулярністю
лінійної ізотерми (LIR).  Запропоноване  узагальнене
рівняння стану може включати в себе LIR рівняння стану  як частковий
випадок. Трипараметрична узагальнена регулярність лінійної ізотерми [Parsafar~G.A., Mason~E.A., Phys. Rev.~B,
1994, \textbf{49},  3049] (PM),  [Shanker~J.,  Singh~B.,  Kushwa~S.S.,  Physica~B, 1997, \textbf{229}, 419] (SSK),
[Parsafar~G.A.,  Spohr~H.V., Patey~G.N., J.~Phys. Chem.~B, 2009,
\textbf{113}, 11980] (PSP) і  переформульовані PM  SSK рівняння стану є застосовані до 30
металічних твердих тіл у широкій області тиску. Показано, що PM, PMR
і PSP рівняння стану для більшості твердих тіл та SSK і SSKR
рівняння стану для декількох твердих тіл мають фізично некоректні
поворотні точки, і тиск стає негативним при досить високому тиску.
Узагальнене  рівняння стану  є здатним не тільки подолати проблему, існуючу
в інших п'яти рівняннях стану, де тиск стає негативним при високому
тиску, але також дає кращі результати, ніж інші рівняння стану.
\keywords трипараметричне рівняння стану, металічні тверді тіла,
високий тиск, фізично некоректні осциляції
\end{abstract}

\end{document}